\documentclass{article}
\usepackage{spconf,amsmath,epsfig}
\usepackage{txfonts}

\newcommand{\mean}{\mathop{\mathrm{mean}}}

\newcommand{\DE}{\mathop{\mathrm{DE}}}
\newcommand{\MF}{\mathop{\mathrm{MF}}}
\newcommand{\MMSE}{\mathop{\mathrm{MMSE}}}

\newcommand{\ZEROvec}{\mathop{\textbf{0}}}

\newcommand{\eye}{\mathop{\mathrm{I}}}

\newcommand{\real}{\varmathbb{R}}

\newcommand{\ba}{\ensuremath{{\mathbf a}}}

\newcommand{\bc}{\ensuremath{{\mathbf c}}}

\newcommand{\bp}{\ensuremath{{\mathbf p}}}

\newcommand{\br}{\ensuremath{{\mathbf r}}}
\newcommand{\bs}{\ensuremath{{\mathbf s}}}

\newcommand{\bw}{\ensuremath{{\mathbf w}}}

\newcommand{\bA}{\ensuremath{{\mathbf A}}}

\newcommand{\bC}{\ensuremath{{\mathbf C}}}

\newcommand{\bS}{\ensuremath{{\mathbf S}}}

\newcommand{\cG}{\ensuremath{{\mathcal G}}}
\newcommand{\cK}{\ensuremath{{\mathcal K}}}

\title{ENERGY EFFICIENCY IN MULTI-HOP CDMA NETWORKS:
A GAME THEORETIC ANALYSIS CONSIDERING OPERATING COSTS}
\name{Sharon Betz and H. Vincent Poor\thanks{This research was supported by the U. S. Air Force Research
Laboratory under Cooperative Agreement
FA8750-06-1-0252, the Defense Advanced Research Projects Agency under
 Grant HR0011-06-1-0052,  the U. S. National Science Foundation under
Grant ANI-03-38807, and an Intel Fellowship.}}
\address{Department of Electrical Engineering\\
Princeton University\\
\{sbetz,poor\}@princeton.edu}
\begin{document}
%
\maketitle
\begin{abstract}
A game-theoretic analysis is used to study the effects of receiver choice and transmit power on the energy efficiency of multi-hop networks in which the nodes communicate using  Direct-Sequence Code Division Multiple Access (DS-CDMA).  A Nash equilibrium of the game  in which the network nodes can choose their receivers as well as their transmit powers to maximize the total number of bits they transmit per unit of energy spent (including both transmit and operating energy) is derived.  The energy efficiencies resulting from the use of different  linear multiuser receivers in this context are compared for the non-cooperative game. Significant gains in energy efficiency are observed when multiuser receivers, particularly  the linear minimum mean-square error (MMSE) receiver, are used instead of conventional matched filter receivers.
\end{abstract}
\section{Introduction}
\label{sec:intro}
In a wireless multi-hop network, nodes communicate by passing
messages for one another; permitting multi-hop communications,
rather than requiring one-hop communications, can increase network capacity and allow for a more ad hoc (and thus scalable) system (with little or no centralized control).  For these reasons, and because of their potential for commercial, military, and civil applications, wireless multi-hop networks have attracted considerable attention over the past few years.  In these networks, energy efficient communication is important because the nodes are typically battery-powered and therefore energy-limited.  Work on energy-efficient communication in these multi-hop networks has often focused on routing protocols; this work instead looks at power control and receiver design choices that can be implemented independently of (and thus in conjunction with) the routing protocol.

One approach that has been very successful in researching energy efficient communications in both cellular and multi-hop networks is the game-theoretic approach described in \cite{GoodmanMandayam00, SaraydarMandayamGoodman02}.  Much of the game-theoretic research in multi-hop networks has focused on pricing schemes (e.g. \cite{IleriMauMandayam05, MarbachQiu05}).  In this work, we avoid the need for such a pricing scheme by using instead a  nodal utility function to capture the energy costs.  It further differs from previous research by considering receiver design, as \cite{Meshkatiet05} does for cellular networks.  This work further differs from existing research, including \cite{BetzPoor06}, through an extension of the utility function that considers the total energy costs, not just the transmit energy.

We propose a distributed noncooperative game in which the nodes can choose their transmit power and linear receiver design to maximize the number of bits that they can send per unit of power.  After describing the network and internodal communications in Section 2, we describe the Nash equilibrium for this game, as well as for a set of games with set receivers, in Section 3.  We present numerical results
and a conclusion in Sections 4 and 5.

\section{System Model}
Consider a wireless multi-hop network with $K$ nodes (users) and an
established logical topology, where a sequence of connected
link-nodes $l\in L(k)$ forms a route originating from a source $k$
(with $k\in L(k)$ by definition).  Let $m(k)$ be the node after node
$k$ in the route for node $k$.  Assume that all routes that go
through a node $k$ continue through $m(k)$ so that node $k$
transmits only to $m(k)$.\footnote{If the routing has a node, $k$, transmitting to multiple nodes, new ``nodes'' can be introduced, collocated with $k$, each with a different destination.  The modifications to the results, taking into account the channel dependence, is straightforward.} Nodes communicate with each other using
DS-CDMA with processing gain $N$ ($N$ chips per bit).

The signal received at any node $m$ (after chip-matched filtering)
sampled at the chip rate over one symbol duration can be expressed
as
\begin{align}
  \br^{(m)} &= \sum_{k=1}^K\sqrt{p_{k}} h_{k}^{(m)} b_k \bs_k + \bw^{(m)}
\end{align}
where $p_{k}$, $b_k$, and $\bs_k$ are the transmit power,
transmitted symbol, and (binary) spreading sequence for node $k$;
$h_{k}^{(m)}$ is the channel gain between nodes $k$ and $m$; and
$\bw^{(m)}$ is the noise vector which is assumed to be Gaussian with
mean $\ZEROvec$ and covariance $\sigma^2\eye$.  (We assume here
$p_m=0$.) Assume the spreading sequences are random, i.e., $\bs_k =
\frac1{\sqrt{N}}[v_1\ldots v_N]^T$, where the $v_i$'s are
independent and identically distributed (i.i.d.) random variables
taking values $\{-1,+1\}$ with equal probabilities.  Denote the
cross-correlations between spreading sequences as $\rho_{kj} = \bs_k^T\bs_j$, noting that $\rho_{kk} = 1$ for all $k$.

Let the vector $\bc_k^{(m)}$ represent the linear receiver at the $m$th node for the $k$th signature sequence.  The
output of this receiver can be written as
\begin{align}
  y &= {\bc_k}^T \br^{(m)}\\
  &= \sqrt{p_{k}}h_{k}^{(m)} b_k {\bc_k}^T\bs_k +
  \sum_{j\neq k}\sqrt{p_{j}} h_{j}^{(m)} b_j {\bc_k}^T\bs_j + {\bc_k}^T\bw^{(m)}.
\end{align}
The signal-to-interference-plus-noise ratio (SINR), $\gamma_k$, of
the $k$th user at the output of receiver $m(k)$ is
\begin{align}\label{SINRkm}
  \gamma_k &= \frac{p_{k}{h_{k}^{(m(k))}}^2 \left({\bc_k}^T\bs_k\right)^2}
  { \sigma^2{\bc_k}^T{\bc_k}+
  \sum_{j\neq k}p_{j}{h_{j}^{(m(k))}}^2 \left({\bc_k}^T\bs_j\right)^2 }.
\end{align}

Each user has a utility function that is the ratio of its effective
throughput to its expended transmit and computation power, i.e.,
\begin{align}\label{util}
  u_k &= \frac{T_k}{p_k+q_k}.
\end{align}
Here, the throughput, $T_k$, is the net number of information bits
sent by $k$ (generated by $k$ or any node whose route goes through
$k$) and received without error at the intended destination, $m(k)$,
per unit of time and $q_k$ is the power expended by the node to
implement the receiver. (We assume that all the congestion control
is done in the choice of routing.)

Following the discussion in \cite{Meshkatiet05}, we will use
\begin{align}\label{Tksimple}
  T_k &= \frac{L}{M}R f(\gamma_k)
\end{align}
where $L$ and $M$ are the number of information bits and the total
number of bits in a packet, respectively (without loss of generality
assumed here to be the same for all users); $R$ is the transmission
rate, which is the ratio of the bandwidth to the processing gain and
is taken for now to be equal for all users;  and $f(\cdot)$ is an
efficiency function that closely approximates the packet success
rate.  This efficiency function can be any increasing, continuously
differentiable, sigmoidal\footnote{A continuous increasing function
is sigmoidal if there is an inflection point above which the function is concave
and below which the function is convex.} function with $f(0)=0$ and
$f(+\infty)=1$.  Let its first derivative be denoted as $f'(\gamma)=\frac{\partial f(\gamma)}{\partial \gamma}$ and let $\gamma_0$ be its inflection point.  See \cite{Meshkatiet05} for more discussion of the
efficiency function.

Using \eqref{Tksimple}, \eqref{util} becomes
\begin{align}\label{utilsimple}
   u_k &= \frac{L}{M}R \frac{f(\gamma_k)}{p_k+q_k}.
\end{align}

When the receiver used is a matched filter (MF) (i.e.
$\bc_k^{(m(k))}=\bs_k$), the received SINR is
\begin{align}\label{SINRMF}
  \gamma_k^{\MF}
  &=
   \frac{p_{k}{h_k^{m(k)}}^2  \left({\bs_k}^T\bs_k\right)^2}
  { \sigma^2{\bs_k}^T{\bs_k}+
  \sum_{j\neq k}p_{j}{h_j^{m(k)}}^2  \left({\bs_k}^T\bs_j\right)^2 }
  \\&=
   \frac{p_{k}{h_k^{m(k)}}^2 }
  { \sigma^2+
  \sum_{j\neq k}p_{j}{h_j^{m(k)}}^2 \rho_{kj}^2 }.
\end{align}

When the receiver is a linear minimum mean-squared error (MMSE)
receiver , the filter coefficients and the received SINR are
\cite{Verdu}
\begin{align}
\label{MMSEcoef}
 \bc_k^{\MMSE} &= \frac{\sqrt{p_k}h_k^{m(k)}}{1+p_k {h_k^{m(k)}}^2 \bs_k^T \bA_k^{-1} \bs_k}\bA_k^{-1} \bs_k
\intertext{and}\label{SINRMMSE}
  \gamma_k^{\MMSE}
  &= p_k {h_k^{m(k)}}^2 \bs_k^T \bA_k^{-1} \bs_k,
\end{align}
where $\bA_k = \sigma^2\eye+\sum_{j\neq k}p_j {h_j^{m(k)}}^2\bs_j\bs_j^T$.

When the receiver is a decorrelator\footnote{Here, we must assume
that $K\leq N$.} (DE) (i.e. $\bC = [\bc_1 \cdots \bc_K] =
\bS(\bS^T\bS)^{-1}$ where $\bS = [\bs_1 \cdots \bs_K]$), the
received SINR is
\begin{align}\label{SINRDE}
  \gamma_k^{\DE} &= \frac{p_k{h_k^{m(k)}}^2}{\sigma^2\bc_k^T\bc_k}.
\end{align}

For any linear receiver with all nodes' coefficients chosen independently of their transmit powers  (including the MF and DE), as well as for the MMSE receiver, the SINR for user $k$ is the product of user $k$'s power and a factor that is independent of user $k$'s power: $\gamma_k(p_k, \bp_{-k}) = p_k g_k(\bp_{-k})$,
where $\bp_{-k}$ is a vector of the powers of all users except for
user $k$ and $g_k$ is a function that depends on the receiver type,
the channel gains, $q_k$, and the users' spreading sequences. This
means that
\begin{align}\label{SINRder}
  \frac{\partial\gamma_k}{\partial p_k} &= \frac{\gamma_k}{p_k} = g_k(\bp_{-k}),
\end{align}
so $\gamma_k$ is strictly increasing in $p_k$.  Thus, for a fixed
receiver type and fixed powers for the other users, there is a
one-to-one relationship between the power of user $k$ and its SINR.
Let $p_0(\bp_{-k}) = \frac{\gamma_0}{g_k(\bp_{-k})}$ be the unique
positive number for which
$\gamma_k^{r}(p_0(\bp_{-k}),\bp_{-k})=\gamma_0$, where, as before,
$\gamma_0$ is the inflection point of the efficiency function
$f(\gamma)$.

\section{The Noncooperative Power-Control Game}

Let $\cG = \Bigl[\cK,\{A_k\},\{u_k\}\Bigr]$ denote the
noncooperative game where $\cK = \{1,\ldots,K\}$ and $A_k =
[0,P_{\max}]\times\mathfrak{R}$ is the strategy set for the $k$th
user. Here, $P_{\max}$ is the maximum allowed power for transmission
and $\mathfrak{R}$ is the set of allowable receivers, for now
restricted to the MF, DE, and MMSE receivers. Each strategy in $A_k$ can be written as
$\ba_k=(p_k,r_k)$, where $p_k$ and $r_k$ are the transmit power and
the receiver type, respectively, of user $k$. Then the resulting
noncooperative game can be expressed as the maximization problem for
$k=1,\ldots,K$:
\begin{align}\label{noncoop}
\max_{\ba_k}u_k &= \max_{p_k,r_k}u_k(p_k,r_k)
\\&=
\frac{L}{M}R\max_{r_k}\left(\max_{p_k}\frac{f(\gamma_k^{r_k}(p_k,\bp_{-k}))}{p_k+q_k^{r_k}}\right),
\end{align}
where $\gamma_k$ and $q_k$ are expressed explicitly as functions of
the transmit power and receiver type.

For each of the receivers, $r$, in $\mathfrak{R}$, let $\cG_{r} =
\Bigl[\cK,\{[0,P_{\max}]\},\{u_k\}\Bigr]$ denote the noncooperative
game that differs from $\cG$ in that users cannot choose their
linear receivers but are forced to use the receiver $r$.  The
resulting noncooperative game can be expressed as the following
maximization problem for $k=1,\ldots,K$:
\begin{align}\label{noncoop2}
\max_{p_k}u_k(p_k,r) =
\frac{LR}{M}\max_{p_k}\frac{f(\gamma_k^{r}(p_k,\bp_{-k}))}{p_k+q_k}.
\end{align}

The following results are summarized without proof due to space constraints.

When given the choice between receivers, it is optimal for the users to use MMSE receivers.  Given a certain system and fixed powers ($\bp$) for all other users, there is a unique optimal power level for each user, $\widetilde{\bp}(\bp)$, that satisfies for each
  $k$
\begin{align}\label{pstar}
  \left.\frac{\partial} {\partial p} \frac{f(\gamma_k(p,\bp_{-k}))} {p+q_k}\right|_{p=\widetilde{p}_k} &= 0
\end{align}
and it occurs in the concave region of the efficiency function:
$\widetilde{p}_k(\bp)>p_0\ \forall\bp\in\real_+^K$.  The game has at least one Nash equilibrium and for any Nash equilibrium, $\bp'$, it holds that $\widetilde{\bp}(\bp')=\bp'$.  Then, as long as $\bp\geq\bp'\implies\widetilde{\bp}(\bp)\geq\widetilde{\bp}(\bp')$ (that is, that a node never lowers its power when other nodes don't lower theirs),
 the Nash equilibrium is unique.  Furthermore, the algorithm where, for each time $t$, the users use
  the power described by
$\bp(t)=\widetilde{\bp}\bigl(\bp(t-1)\bigr)$ converges to the unique
Nash equilibrium for any initial choice of power vector.

The solution to $\gamma f'(\gamma)=f(\gamma)$ is a lower bound on the achieved SINR at the Nash equilibrium; as $q_k$ increases, so does $\gamma_k$.  That is, as the power necessary to run increases, the nodes aim for a higher SINR: to make the transmission worthwhile, they need more throughput.  Because of this, for the MF and MMSE receivers, the utilities of \emph{all} users decreases when any user's value of $q_k$ increases (for the DE receiver, only the $k$th user's utility decreases).

  For the decorrelator, the Nash equilibrium is Pareto optimal.  For the MMSE and MF receivers,  the Nash equilibrium is not Pareto optimal and can be improved upon if every user decreases its power by a small factor.

\section{Numerical Results}

  Consider a multi-hop network with $K$ nodes distributed randomly in a square whose area is $100K$ square km, surrounding an access point in the center.  For simplicity, the simulations assume a routing scheme where all nodes transmit to the closest node that is closer to the access point (or the access point of that is closest).  The packets each contain 100 bits of data and no overhead ($L=M=100$); the transmission rate is $R=100$ kb/s; the thermal noise power is $\sigma^2=5\times 10^{-16}$ Watts;  the channel gains are distributed with a Rayleigh distribution with mean $0.3d^{-2}$, where $d$ is the distance between the transmitter and receiver; and the processing gain is $N=32$.  We use the same efficiency function as \cite{Meshkatiet05}, namely $f(\gamma)=(1-e^{-\gamma})^M$, which can be shown to satisfy the conditions for the existence of a unique Nash equilibrium.  Finally, the amount of energy that a node has to expend to run, $q_k$, is assumed to be the same for all nodes and is allow to range from $0.0001$ Joules to $1$ Joule per transmission (equivalently, for the rate and packet size given, it ranges between $0.001$ Watts and 10 Watts).

\begin{figure}
\begin{centering}
  \includegraphics[width=3.5in]{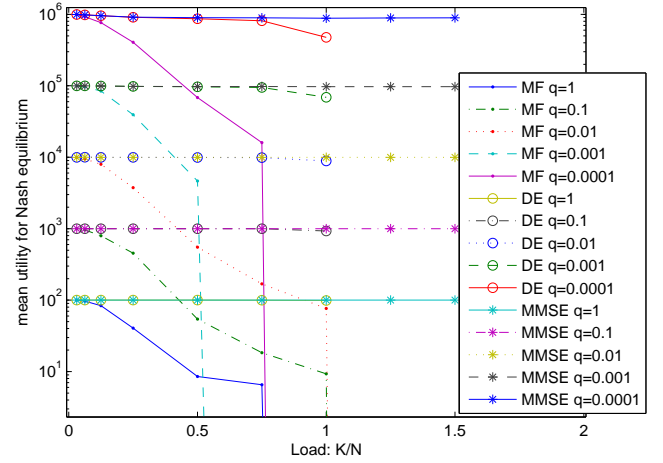}\\
  \caption{Mean utility for the different receivers}\label{meanutil_beta}
\end{centering}
\end{figure}

Figure \ref{meanutil_beta} shows the mean utility (averaged over 100 realizations of the system) for the different receivers as a function of the system load.  Here the load ranges from 0 to 1.5.  For loads greater than 1, the DE receiver cannot be used. Changing the value of $q$ by a factor of ten for all the users changes the mean utility by roughly a factor of ten as well.  The MF receiver is most affected by the increased load, with the mean utility dropping by about a factor of ten when the load increases from 10\% to 50\%.  The performance of the DE and MMSE receivers are similar, although the MMSE receiver outperforms the DE receiver at all points, with more significant gains at high load.

\begin{figure}
\begin{centering}
  \includegraphics[width=3.2in]{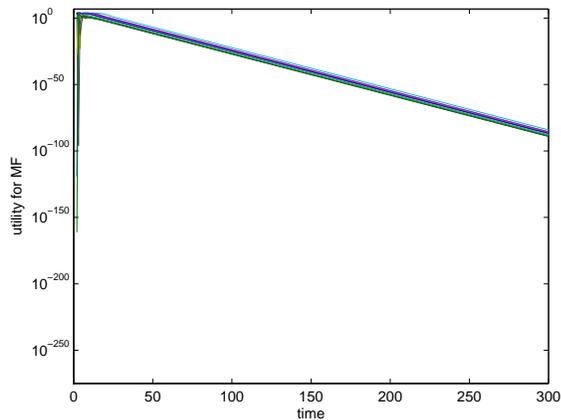}\\
  \caption{Utility for all 16 users using the MF for one scenario with $\beta=\frac12$ and $q=0.01$}\label{MFutil_time}
\end{centering}
\end{figure}

\begin{figure}
\begin{centering}
  \includegraphics[width=3.2in]{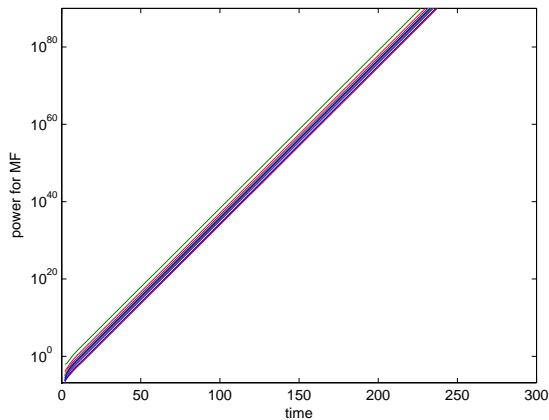}\\
  \caption{Power for all 16 users using the MF for one scenario with $\beta=\frac12$ and $q=0.01$}\label{MFP_time}
\end{centering}
\end{figure}

As shown above, for the MMSE and MF receivers, the Nash equilibrium point is not Pareto optimal.  This is particularly obvious for the matched filter.  Figure \ref{MFutil_time} shows how the utility for all 16 users in one scenario change with time when the simple algorithm described above is run.  Notice that all users have better utility before convergence.  In fact, for this example, $\mean\{u_k\}$ achieves its maximum at $t=2$ while $\min_k\{u_k\}$ achieves its maximum at $t=6$.  Figure \ref{MFP_time} shows the transmit power for all users in the same example; from this it is clear that the users are transmitting at higher and higher powers to the detriment of their utilities.  The MMSE receiver tends to do much better (though it would still be better for all users to use slightly less power each); the MMSE receiver also converges much faster.  The DE receiver converges in just one time step (due to the independence of the power choices of different users) and results in a Pareto optimal solution.

\section{Conclusion}
We have analyzed the cross-layer issue of energy efficient communication in multi-hop networks using a game theoretic model.  We've extended previous work in this area to consider the energy costs used in running the receiver and transmitter, in addition to the actual transmit costs.  Amongst all linear receivers, the MMSE receiver is optimal.  For the MF, MMSE, and DE receivers, a unique Nash equilibrium exists, though for the MF and MMSE receivers, this Nash equilibrium is not Pareto optimal.

\bibliographystyle{IEEEtran}

\end{document}